\documentclass[aps,preprint,onecolumn,nofootinbib]{revtex4}
\usepackage{dcolumn}
\usepackage{bm}
\usepackage{mathrsfs}
\usepackage{amssymb}
\usepackage{amsmath}
\usepackage{amsfonts}
\usepackage{color}
\usepackage{ulem}

\usepackage[]{graphicx}
\begin{document}
\bibliographystyle{prsty}
\title{An Elementary Proof That Everett's Quantum Multiverse Is Nonlocal: Bell-Locality and Branch-Symmetry in the Many-Worlds Interpretation}
\author{ Aur\'elien Drezet$^{1}$}
\address{(1) Institut NEEL, CNRS and Universit\'e Grenoble Alpes, F-38000 Grenoble, France }
\email{aurelien.drezet@neel.cnrs.fr}
\begin{abstract}
 {Everett's many-worlds or multiverse} 
 theory is an attempt to find an alternative to the standard Copenhagen interpretation of quantum mechanics. Everett's theory is often  claimed to be local in the Bell sense. Here, we show that this is not the case and debunk the contradictions by analyzing in detail the Greenberger--Horne--Zeilinger (GHZ) nonlocality theorem. We discuss and compare different notions of locality often mixed in the Everettian literature and try to explain the nature of the confusion. We conclude with a discussion  of probability and statistics in the many-worlds theory and stress that the strong symmetry existing between branches in the theory prohibits the definition of probability and that the theory cannot recover statistics.    The only way out from this contradiction is to modify the theory by adding hidden variables \`a la Bohm and, as a consequence, the new theory is explicitly Bell-nonlocal.    
\end{abstract}

\maketitle
\section{Goal of the Work}
\label{Sec1}
\indent It has often been repeated that Everett's interpretation~\cite{Barrett2012,Everett1957} (the so called many-worlds interpretation) of quantum mechanics is local in the Einstein or Bell sense of the word. The claim goes back to Everett's work~\cite{Barrett2012,Everett1957} where he mentioned that his theory solves the  Einstein--Podolsky--Rosen (EPR) contradiction \cite{EPR} in a local way.  In the same vein, locality claims involving Bell's theorem \cite{Bell} and Greenberger--Horne--Zeilinger (GHZ) states \cite{GHZ} have been discussed by Page \cite{Page1982}, Blaylock~\cite{Blaylock}, Vaidman~\cite{Vaidman1994,Vaidman2015}, Tipler \cite{Tipler2014}, Timpson and Brown~\cite{Timpson2003,Timpson2015}, and many others \cite{Price1995,Sakaguchi1996,Deutsch1999,Rubin2001,Horsman,Brassard,Bedard2021,Waegell,WallaceBook}, including popular science books~\cite{Bruce,Carroll}. Unfortunately, these claims are based on a misunderstanding of EPR and Bell works~\cite{EPR,Bell}, and it is the purpose of this article to explain why it is so.\\
\indent The organization of the article is as follows. In Section \ref{Sec2}, we review and discuss the GHZ theorem and emphasize the roles of beables and local-causality. In Section \ref{Sec3}, we analyze and debunk the claims for locality in Everett's theory.  In the {Section} 
 \ref{Sec4}, we discuss the previous results in connection with the `incoherence problem' associated with the absence of probability in Everett's theory. We stress that the strong symmetry between the different branches in the many-worlds theory prohibits the existence of an unambiguous definition of probability and statistics in this theory.  The only way out from this contradiction is to modify the ontology  by adding supplementary  hidden variables as in the de~Broglie--Bohm theory~\cite{Bohm}. However, the new theory~\cite{Valentini2010,Tipler2014,Bostrom2014,Sebens2014,Hall2014,Tappenden} naturally displays action-at-a-distance, i.e.,~is Bell-nonlocal. 
\section{The GHZ Nonlocality Theorem}
\label{Sec2}
\indent We use the GHZ nonlocality proof~\cite{GHZ} considered by Vaidman~\cite{Vaidman2015}, which has the great advantage (especially in the context of Everett's theory) of not requiring  probabilistic 
Bell-like inequalities (more on that in Section \ref{Sec4}). The GHZ state of three spin-$\frac{1}{2}$ maximallly entangled particles {reads:} 
\begin{eqnarray}
|\psi\rangle=\frac{1}{\sqrt{2}}(|+_z\rangle^1|+_z\rangle^2|+_z\rangle^3-|-_z\rangle^1|-_z\rangle^2|-_z\rangle^3),\label{GHZ}
\end{eqnarray} where $|\pm_z\rangle^j$ are the eigenstates of the $\sigma_z^{(j)}$ Pauli matrix for particle $j$. Remarkably, we have:   
\begin{eqnarray}
\sigma_x^{(1)}\sigma_x^{(2)}\sigma_x^{(3)}|\psi\rangle=-|\psi\rangle\\
\sigma_x^{(1)}\sigma_y^{(2)}\sigma_y^{(3)}|\psi\rangle=+|\psi\rangle\\
\sigma_y^{(1)}\sigma_x^{(2)}\sigma_y^{(3)}|\psi\rangle=+|\psi\rangle\\
\sigma_y^{(1)}\sigma_y^{(2)}\sigma_x^{(3)}|\psi\rangle=+|\psi\rangle.
\end{eqnarray} 

EPR and Bell deductions start with an Einstein locality assumption, ensuring that local measurements made by distance observers Alice, Bob, and Charlie on particles 1, 2, and 3 are independent and not influenced by each other. Locality applied to the previous GHZ system allows us to use a counterfactual reasoning in order to define elements of reality $v(\sigma_x^{(j)})=\pm 1$ and   $v(\sigma_y^{(j)})=\pm 1$. More precisely, we have the four counterfactual conditions:  
\begin{eqnarray}
v(\sigma_x^{(1)})v(\sigma_x^{(2)})v(\sigma_x^{(3)})=-1\\
v(\sigma_x^{(1)})v(\sigma_y^{(2)})v(\sigma_y^{(3)})=+1\\
v(\sigma_y^{(1)})v(\sigma_x^{(2)})v(\sigma_y^{(3)})=+1\\
v(\sigma_y^{(1)})v(\sigma_y^{(2)})v(\sigma_x^{(3)})=+1,
\end{eqnarray} which have to be simulteanously fulfilled in order to agree with the locality assumption. Moreover, Equation (6) conflicts with the product of Equations (7)--(9), leading to   $v(\sigma_x^{(1)})v(\sigma_x^{(2)})v(\sigma_x^{(3)})=+1$. Therefore, the locality condition conflicts with quantum mechanics; quantum mechanics is nonlocal.\\
\indent It is important to stress that standard quantum mechanics, like most realist interpretations (e.g., involving hidden variables), assumes that each observer can only observe one outcome during the experiment.  This single outcome corresponds to  one of the possible  values of the different elements of reality  $v(\sigma_x^{(j)})$, $v(\sigma_y^{(j)})$.  This looks like an additional assumption, and it is known that the many-worlds interpretation does not  require  such an assumption. Therefore,   an Everettian would claim  that the GHZ theorem cannot be applied if the single outcome rule is not assumed. We will reply to this repeated claim in Section \ref{Sec3}, but as a spoiler, we can already state that the only thing that is needed  in order to apply the GHZ theorem to the many-worlds interpretation is the confirmation that each observer only observe a single outcome {\em in his/her relative 
 branch}.  \\       
\indent It is also important to stress that the counterfactual existence of elements of reality is actually deduced from locality. More exactly, and contrarily to the `common wisdom' ({which 
  refers to claims often presented in the literature; for more on this issue, the readers could consider the articles by Maudlin published in 2014~\cite{Maudlin2014} and the discussion with Werner \cite{Werner2014} and others \cite{Zukowski2014}; see also \cite{Laudisa2023,Drezetarxiv})}, EPR, Bell, and GHZ theorems do not require realism as an independent assumption. Bell--GHZ theorems are not theorems about local-realism but only about locality. Bell was very clear about that when he introduced the concept of beables $\lambda$ \cite{Bell}. In the GHZ scenario (as in the EPR--Bell scenario involving a spin singlet), the perfect correlations and locality impose determinism ({in 
 all these ideal examples, we neglect losses and absorption involving imperfect detectors}),  and we have   $v(\sigma_x^{(j)})=A_j(\lambda)$, $v(\sigma_y^{(j)})=B_j(\lambda)$, meaning that the elements of reality are deterministic functions of the beables' $\lambda$-defining causes located in the past of the system. That issue is already true for the EPR--Bell scenario (neglecting the detection loophole), and we have, in both cases, quantum mechanics local $\Rightarrow$ quantum mechanics deterministic. From this, we naturally get quantum mechanics deterministic $\Rightarrow$ quantum mechanics incomplete, which is the EPR dilemma: quantum mechanics is either incomplete or nonlocal.\\ 
\indent Crucially, this determinism implies the joint validity of the four conditions in Equations~(6)--(9) that can be rewritten as:
\begin{eqnarray}
A_1(\lambda)A_2(\lambda)A_3(\lambda)=-1\\
A_1(\lambda)B_2(\lambda)B_3(\lambda)=+1\\
B_1(\lambda)A_2(\lambda)B_3(\lambda)=+1\\
B_1(\lambda)B_2(\lambda)A_3(\lambda)=+1,
\end{eqnarray} leading to the GHZ contradiction and thus to the conclusion that quantum mechanics must be nonlocal.\\
\indent This is already enough to qualitatively understand the GHZ theorem. Moreover, in order to appreciate its generality and how it applies to Everett's theory, it is central here to discuss in more detail the concept of beables and local-causality. We thus decompose the whole deduction in several steps, the first of which being the introduction of beables. The crux is that $\lambda$-defining causes do not necessarily involve hidden variables (which would be introduced in order to complete quantum mechanics). $\lambda$ can also include purely quantum variables associated with the wave function $|\psi\rangle$. To better understand, we introduce with Bell the probability for joint observables, e.g., $\mathcal{P}(\alpha,\beta,\gamma|\hat{\mathbf{n}}_1,\hat{\mathbf{n}}_2,\hat{\mathbf{n}}_3,\psi)$ [with $\alpha,\beta,\gamma$ the measurement outcomes  with values $\pm 1$, and $\hat{\mathbf{n}}_1,\hat{\mathbf{n}}_2,\hat{\mathbf{n}}_3$ the spin directions (unit vectors) of the three independent Stern--Gerlach analyzers of Alice, Bob, and Charlie], and using beables we~write: 
\begin{eqnarray}
\mathcal{P}(\alpha,\beta,\gamma|\hat{\mathbf{n}}_1,\hat{\mathbf{n}}_2,\hat{\mathbf{n}}_3,\psi)=\int_\Lambda d\lambda\mathcal{P}(\alpha,\beta,\gamma|\lambda,\hat{\mathbf{n}}_1,\hat{\mathbf{n}}_2,\hat{\mathbf{n}}_3,\psi)\rho(\lambda|\hat{\mathbf{n}}_1,\hat{\mathbf{n}}_2,\hat{\mathbf{n}}_3,\psi).
\end{eqnarray} 

This relation is  general at that stage and, in particular, it is true even as we assume quantum mechanics to be a complete theory. For instance, according to Beltrametti and Bugajski~\mbox{\cite{Beltrametti1995,Drezet2012}}, for a complete theory, we can always write: 
\begin{eqnarray}
\mathcal{P}(\alpha,\beta,\gamma|\hat{\mathbf{n}}_1,\hat{\mathbf{n}}_2,\hat{\mathbf{n}}_3,\psi)
=\int_{\Theta}  \mathcal{P}(\alpha,\beta,\gamma|\hat{\mathbf{n}}_1,\hat{\mathbf{n}}_2,\hat{\mathbf{n}}_3,\theta)\delta(\theta-\psi)d\theta,\label{beltra}
\end{eqnarray}  where now the beable $\theta\equiv |\theta\rangle$ belongs to the Hilbert space of the two-spins system. We~have:
\begin{eqnarray}
\mathcal{P}(\alpha,\beta,\gamma|\theta,\hat{\mathbf{n}}_1,\hat{\mathbf{n}}_2,\hat{\mathbf{n}}_3)=|\langle\alpha_{\hat{\mathbf{n}}_1},\beta_{\hat{\mathbf{n}}_2},\gamma_{\hat{\mathbf{n}}_3}|\theta\rangle|^2
\end{eqnarray} and $\rho(\theta |\psi)\equiv\delta(|\theta\rangle-|\psi\rangle)$. {More 
 rigorously writing $|\theta\rangle=\sum_{\alpha,\beta,\gamma}\theta_{\alpha,\beta,\gamma}| \alpha_{\hat{\mathbf{n}}_1},\beta_{\hat{\mathbf{n}}_2}, \gamma_{\hat{\mathbf{n}}_3} \rangle$ with $\theta_{\alpha,\beta,\gamma}:=\langle\alpha_{\hat{\mathbf{n}}_1},\beta_{\hat{\mathbf{n}}_2},\gamma_{\hat{\mathbf{n}}_3}|\theta\rangle$ $\in \mathbb{C}$, a generally complex valued amplitude (with $\alpha,\beta,\gamma=\pm 1$), we have $\delta(|\theta\rangle-|\psi\rangle):=\prod_{\alpha,\beta,\gamma}\delta(\theta'_{\alpha,\beta,\gamma}-\psi'_{\alpha,\beta,\gamma})\delta(\theta''_{\alpha,\beta,\gamma}-\psi''_{\alpha,\beta,\gamma})$ with $\psi_{\alpha,\beta,\gamma}:=\langle\alpha_{\hat{\mathbf{n}}_1},\beta_{\hat{\mathbf{n}}_2},\gamma_{\hat{\mathbf{n}}_3}|\psi\rangle$,  and $d|\theta\rangle:=\prod_{\alpha,\beta,\gamma}d\theta'_{\alpha,\beta,\gamma}d\theta''_{\alpha,\beta,\gamma}$ ($\theta'_{\alpha,\beta,\gamma}$, and $\theta''_{\alpha,\beta,\gamma}$ are the real and imaginary parts of $\theta_{\alpha,\beta,\gamma}$ and similarly for $\psi_{\alpha,\beta,\gamma}$). We also have $\mathcal{P}(\alpha,\beta,\gamma|\theta,\hat{\mathbf{n}}_1,\hat{\mathbf{n}}_2,\hat{\mathbf{n}}_3)=|\theta_{\alpha,\beta,\gamma}|^2$.}  This representation corresponds to what Harrigan and Spekkens \cite{Harrigan2010} call an ontic description of the quantum state where the Dirac distribution associates a strongly localized density of probability to $|\psi\rangle$. It is thus wrong to assume  that $\lambda$ necessarily denotes an hidden variable. Unfortunately, this claim has been too often repeated, resulting in great misunderstanding concerning the goal and generality of Bell and GHZ theorems.\\
\indent The next step in the Bell--GHZ deduction involves the local-causality condition \cite{Bell,Bell2023}, which assumes:
  \begin{eqnarray}
\mathcal{P}(\alpha,\beta,\gamma|\lambda,\hat{\mathbf{n}}_1,\hat{\mathbf{n}}_2,\hat{\mathbf{n}}_3,\psi)=\mathcal{P}_1(\alpha|\lambda,\hat{\mathbf{n}}_1,\psi)\mathcal{P}_2(\beta|\lambda,\hat{\mathbf{n}}_2,\psi)\mathcal{P}_3(\gamma|\lambda,\hat{\mathbf{n}}_3,\psi) \label{cond1}\\
\rho(\lambda|\hat{\mathbf{n}}_1,\hat{\mathbf{n}}_2,\hat{\mathbf{n}}_3,\psi)=\rho(\lambda|\psi).\label{cond2}
 \end{eqnarray} 
 
 The first factorizability condition in Equation~\eqref{cond1} can be justified under other physical assumptions involving parameter and outcome independences, while the second condition Equation~\eqref{cond2} (also sometimes called the `freedom of choice' condition) involves statistical independence of the density of probability and rejects superdeterminism. We emphasize that most of the time in the literature \cite{Bell2023,Maudlin2014,Norsen2011}, Equation~\eqref{cond2} is not defined as a part of local-causality and is seen as an independent (and perhaps obvious) condition.  While this is not mandatory, we regroup here \cite{Drezetarxiv} for simplicity  Equations~\eqref{cond1} and \eqref{cond2} in the definition of Bell-locality mixing of both causality and locality (this is further discussed in Appendix \ref{Ap1}). It is important to note that, in the Bell-locality definition, all the beables $\lambda,\hat{\mathbf{n}}_j$ are defined in the past light cones of events associated with beables $\alpha,\beta,\gamma$. This is the core of Einstein's relativistic locality and causality as defined and used by Bell. We stress that Einstein's  and Bell's local-causality must be physically distinguished from local-commutativity and microcausality used in quantum fields theory in order to derive  the non-signaling theorem. This is a central issue that will be discussed in Section \ref{Sec3}. It is also important to stress that  Equation~\eqref{cond2} is valid for the Beltrametti and Bugajski model~\cite{Beltrametti1995,Drezet2012} but that Equation~\eqref{cond1} is not. This already shows that local-causality conflicts with quantum mechanics that is assumed to be complete. \\
 \indent The following step is technical and derives determinism. From Equation~\eqref{cond1}, we define local observable  values such as: \begin{eqnarray}
 A_j(\lambda)=\sum_{\alpha=\pm 1}\alpha\mathcal{P}_j(\alpha|\lambda,\hat{\mathbf{x}},\psi)\nonumber\\
 B_j(\lambda)=\sum_{\alpha=\pm 1}\alpha\mathcal{P}_j(\alpha|\lambda,\hat{\mathbf{y}},\psi)
 \end{eqnarray} where, in general, $|A_j(\lambda)|\leq 1$, $|B_j(\lambda)|\leq 1$ for a generally stochastic theory. Due to probability conservation, we have $\sum_{\alpha=\pm 1}\mathcal{P}_j(\alpha|\lambda,\hat{\mathbf{x}},\psi)=1$). Importantly, the specific  GHZ state given by Equation~\eqref{GHZ} restrains the admissible values  of $A_j(\lambda)$ and $B_j(\lambda)$ to $\pm 1$. Indeed, consider as an example that Alice, Bob, and Charlie measure the spin along the $\hat{\mathbf{x}}$ direction and obtain for the three recorded particles the results  $|+_x\rangle^1|-_x\rangle^2|+_x\rangle^3$ as allowed from Equation~(2). For the actualized $\lambda$ predermining the observation, we of course have $\rho(\lambda|\psi)\neq 0$ and also \begin{eqnarray}
 \mathcal{P}_1(+1|\lambda,\hat{\mathbf{x}},\psi)\neq 0, & \mathcal{P}_2(-|\lambda,\hat{\mathbf{x}},\psi)\neq 0, &\mathcal{P}_3(+1|\lambda,\hat{\mathbf{x}},\psi)\neq 0.
\end{eqnarray}

From Equation~(2),  we must also have, for the same $\lambda$:
\begin{eqnarray}
\mathcal{P}_1(+|\lambda,\hat{\mathbf{n}}_1,\psi)\mathcal{P}_2(-1|\lambda,\hat{\mathbf{n}}_2,\psi)\mathcal{P}_3(-1|\lambda,\hat{\mathbf{n}}_3,\psi)=0\\
 \mathcal{P}_1(-|\lambda,\hat{\mathbf{n}}_1,\psi)\mathcal{P}_2(-1|\lambda,\hat{\mathbf{n}}_2,\psi)\mathcal{P}_3(+1|\lambda,\hat{\mathbf{n}}_3,\psi)=0\\
 \mathcal{P}_1(+|\lambda,\hat{\mathbf{n}}_1,\psi)\mathcal{P}_2(+1|\lambda,\hat{\mathbf{n}}_2,\psi)\mathcal{P}_3(+1|\lambda,\hat{\mathbf{n}}_3,\psi)=0\\
 \mathcal{P}_1(-|\lambda,\hat{\mathbf{n}}_1,\psi)\mathcal{P}_2(+1|\lambda,\hat{\mathbf{n}}_2,\psi)\mathcal{P}_3(-1|\lambda,\hat{\mathbf{n}}_3,\psi)=0.
\end{eqnarray}

Comparing Equation~(20) to Equations (21)--(23), we deduce:
 \begin{eqnarray}
 \mathcal{P}_3(-1|\lambda,\hat{\mathbf{x}},\psi)=\mathcal{P}_1(-|\lambda,\hat{\mathbf{x}},\psi)=\mathcal{P}_2(+1|\lambda,\hat{\mathbf{x}},\psi)=0\\
 \mathcal{P}_3(+1|\lambda,\hat{\mathbf{x}},\psi)=\mathcal{P}_1(+|\lambda,\hat{\mathbf{x}},\psi)=\mathcal{P}_2(-1|\lambda,\hat{\mathbf{x}},\psi)=1,
\end{eqnarray} where Equation~(26) is deduced from Equation~(25) from probability conservation. {Equation~
 (24) is automatically satisfied}. In other words, we deduced that $\mathcal{P}_j(\alpha|\lambda,\hat{\mathbf{x}},\psi)$ can only take the values 0 or 1. Similar results are easily derived from $\mathcal{P}_j(\alpha|\lambda,\hat{\mathbf{y}},\psi)$. This defines a deterministic theory where $A_j(\lambda)$ and $B_j(\lambda)$ are restricted to $\pm 1$ and where we also have 
\begin{eqnarray}
 \mathcal{P}_j(\alpha|\lambda,\hat{\mathbf{x}},\psi)=\delta_{\alpha,A_j(\lambda)}, &\mathcal{P}_j(\alpha|\lambda,\hat{\mathbf{y}},\psi)=\delta_{\alpha,B_j(\lambda)}
\end{eqnarray} together with  the GHZ constraints of Equations~(10)--(13).\\
\indent To summarize our previous analysis, we showed: (i) The introduction of beables $\lambda$ is very general and does not presume the existence of hidden variables. (ii) After recalling the precise definition of Bell-locality (i.e., local-causality) as a compound  condition, we showed  that, for the GHZ state, locality added to quantum mechanics implies determinism. (iii) Finally, determinism with the conditions of Equations~(10)--(13) implies a contradiction, leading to the robust conclusion that quantum mechanics must be nonlocal irrespectively of being complete or not.
\section{Everett and Nonlocality}\label{Sec3}
\indent The question that we want to answer now is why advocates of Everett's theory strongly believe that the Bell--GHZ theorem does not prove nonlocality in their approach?  We can find at least two arguments or reasons in the literature   \cite{Page1982,Blaylock,Vaidman1994,Vaidman2015,Tipler2014,Timpson2003,Timpson2015,Price1995,Sakaguchi1996,Deutsch1999,Rubin2001,Horsman,Brassard,Bedard2021,Waegell,WallaceBook}.\\
\indent \textbf{(A) 
} First is the very intuitive and naive fact that Everett's theory is a non-collapse quantum theory.  In the Copenhagen interpretation, involving an instantaneous wave-function collapse \`a la Heisenberg--von~Neumann, there is clearly a nonlocal feature violating the spirit of Einstein's relativity. In Everett's theory, there is no collapse at all, so we can a~priori hope to dissolve the issue of nonlocality.  The problem with this kind of remark is that, in actuality, the issue is not specific to Everett's attempt. In the de~Broglie--Bohm pilot-wave theory \cite{Bohm}, we also do not have collapse of the wave-function, but the theory is strongly nonlocal and involves instantaneous action-at-a-distance in agreement with the Bell and GHZ theorems. This intuitive argument for locality is thus not very convincing.\\ 
\indent  Moreover, for Everettians, this informal argument is seen as merely a kind of `mise en bouche'. Indeed, there is a  different, more sophisticated  version  of the same reasoning,  which is even more often given by proponents of locality in Everett's theory. The point is that Everett's theory relaxes the `single/unique outcome assumption', which is claimed to be an hidden assumption of Bell--GHZ theorems \cite{Timpson2003,Timpson2015,Horsman,WallaceBook,Tipler2014,Vaidman2015,Waegell}. Since there is more than one outcome existing in different `parallel' independent worlds (contributing to the whole universal wave-function), Bell--GHZ theorems  based on beables $\lambda$ do not apply, and there is no sense in speaking about Bell's nonlocality in Everett's theory---the notion cannot even be clearly formulated. The  claim of Everettians here is that $\lambda$, the beable or cause used in Bell and GHZ  derivations, involves the assumption that a single actualized world corresponding to $\lambda$ has been preselected (i.e., as an hidden variable). However, this contradicts the very basis of Everett's theory, presuming that such a single world does not exist and that there is no need for a hidden variable. Therefore, they claim, Everett's theory cannot be said to be nonlocal in the Bell sense.\\
\indent The problem with this move is that it is based on a sloppy language concerning nonlocality \`a la Bell.  Bell locality is really local-causality and involves both locality and causality. However,  as we saw in Section \ref{Sec2}, this notion of local-causality can be defined independently of any discussion about completeness or incompleteness in quantum mechanics. As we saw with the Beltrametti and Bugajski representation~\cite{Beltrametti1995,Drezet2012},  $\lambda:=\theta$, i.e., the beables, denote causes that can be defined even if quantum mechanics is supposed to be complete \cite{Drezetarxiv}. If we  look at  Equation~\eqref{beltra}, we see that the presence of the Dirac distribution $\delta(\theta-\psi)$ picks up a particular value of the beable $\theta$ corresponding to the wave function $\psi$. This value is defining a cause located in the past of the measurements  made by Alice, Bob, and Charlie. Moreover, since Everett's theory assumes no hidden variable (the theory is `complete' in that sense), the causal representation of Beltrametti and Bugajski in terms of beables $\theta$ can be applied unambiguously to this quantum interpretation: it defines a causal link for the many-worlds theory as well!\\
\indent Moreover, in the many-worlds interpretation, the wave function splits into branches containing observers recording well-defined values  $v(\sigma_x^{(j)})=\pm 1$, $v(\sigma_y^{(j)})=\pm 1$.   If  friends are coming to see Alice, Bob, and Charlie and ask them what eigenvalues they obtained,  they will always give a definite  answer. From the point of view of each separate  branch,  the reality is of course obeying the single outcome rule as in standard quantum mechanics.  This agreement with quantum mechanics  is clearly required in order to apply the conclusions of the GHZ theorem: it is not an independent postulate. What is missed by the Everettians, however, is  that Bell--GHZ theorems prove the impossiblity of building a  beable representation of quantum theory satisfying Equations~\eqref{cond1} and \eqref{cond2} independently on hypotheses concerning completeness or incompleteness.  Everett's theory does not escape  this proof by the introduction of a new and hidden `single outcome' assumption that would be ruled out in the many-worlds theory. A single outcome is not an additional assumption of the Bell--GHZ theorem. What are needed concern: (i) the relative existence of a single outcome {\em in each 
 branch}  (in agreement with quantum mechanics), and (ii) the fact that we can speak about causality using beables $\lambda$, which conflicts with Bell-locality. Furthermore, the GHZ theorem starts with the possibility  of defining beables $\lambda$, which is always true for quantum mechanics! In other words, while a causal  link  involving beables can always be defined  in quantum mechanics (e.g., Equation~\eqref{beltra}), it is  impossible to define a locally-causal relationship for quantum mechanics and a fortiori for the many-worlds interpretation: Everett's theory is definitely nonlocally-causal, i.e., Bell nonlocal.  We point out that the concept of local beables in the many-worlds interpretation is not discussed much in the literature. We provide a short analysis of this issue in Appendix \ref{Ap0} in relation to work \cite{Allori} concerning the ontology of Everett's theory. \\  
\indent \textbf{(B) 
} Perhaps Everettians would resist my conclusion that their theory is Bell nonlocal. They could probably try to escape using a different but related argumentation in favor of locality.  Indeed, there is a second argument based on a strong reliance on locality but on a locality that is actually not the one defined by EPR and Bell.  What the advocates of such a move tacitly call locality is actually the locality used in quantum fields theory involving microcausality and local-commutativity. Local-commutativity means that we have $[\mathcal{O}_A,\mathcal{O}_B]=0$ for two local Hermitian operators $\mathcal{O}_A$ and $\mathcal{O}_B$ defined in two space-like separated spatial regions.   This form of locality allows us to derive the famous non-signaling theorem that in turn implies a `peaceful' relationship between quantum mechanics and special relativity.  More precisely, take as an example the case where Alice is considering the evolution of the mean value $\langle\mathcal{O}_A(t)\rangle$ between times $t$ and $t+\delta t$ when Bob disturbs his settings locally. Assuming (in the interaction picture)   $\mathcal{O}_A(t+\delta t)=\mathcal{O}_A(t)$, and that the quantum state evolves as $|\Psi(t+\delta t)\rangle=e^{-i\delta t\mathcal{O}_B(t)}|\Psi(t)\rangle$,  we obtain, if  $[\mathcal{O}_A,\mathcal{O}_B]=0$:
\begin{eqnarray}
\langle\mathcal{O}_A(t+\delta t)\rangle= \langle \Psi(t)|e^{+i\delta t\mathcal{O}_B(t)}\mathcal{O}_A(t)e^{-i\delta t\mathcal{O}_B(t)}|\Psi(t)\rangle=\langle\mathcal{O}_A(t)\rangle.
\end{eqnarray} 

This condition shows that a local measurement made by Bob on his side cannot affect statistical observables of Alice. This is used to demonstrate the nonsignaling theorem, which is a key feature of relativistic quantum mechanics and quantum information processing. In particular, in the case of the GHZ state, it leads to the nonsignaling condition:
\begin{eqnarray}
\sum_{\beta,\gamma}\mathcal{P}(\alpha,\beta,\gamma|\hat{\mathbf{n}}_1,\hat{\mathbf{n}}_2,\hat{\mathbf{n}}_3,\psi):=\mathcal{P}(\alpha|\hat{\mathbf{n}}_1,\hat{\mathbf{n}}_2,\hat{\mathbf{n}}_3,\psi)=\mathcal{P}_1(\alpha|\hat{\mathbf{n}}_1,\psi).
\end{eqnarray} 

Clearly, local-causality implies nonsignaling, but the opposite is not always true. The advocates of Everett's theory actually define locality using the non-signaling theorem, but this theorem is not specific to Everett's theory and cannot be used to prove anything concerning local-causality \`a la Bell.  Nevertheless, the advocates of Everett's  theory like this weaker form of locality a lot based on non-signaling because it  apparently corresponds with the physical facts each observer on his/her side or in his/her `world' is only aware of: i.e.,~a local ({here 
 meaning `localized',  with a geometrical origin}) piece of information.   The often repeated claim goes like this~\cite{Price1995,Sakaguchi1996,Tipler2014,Blaylock,Timpson2003,Timpson2015,Horsman}:  since Alice, Bob, and Charlie, before communicating and sharing their outcomes, are unaware of the other agents' results, it is unphysical to discuss the counterfactual existence of other observables that are not predicted by the theory. Indeed, in quantum mechanics, as repeated many times by Bohr, the whole experimental  configuration must be taken into consideration, and counterfactual deductions are  generally unsound. Consider, for example, that Alice, Bob, and Charlie actually perform a measurement with $\hat{\mathbf{n}}_1=\hat{\mathbf{n}}_2=\hat{\mathbf{n}}_3=\hat{\mathbf{x}}$. We write the initial GHZ state Equation~1 before the measurement as $|\psi\rangle=\sum_{\alpha,\beta,\gamma} \mathcal{A}_{\alpha,\beta,\gamma}| \alpha_x\rangle^1|\beta_x\rangle^2|\gamma_x \rangle^3$ with $|\mathcal{A}_{\alpha,\beta,\gamma}|^2=1/4$ iff $\alpha\beta\gamma=-1$ (see Equation~(6)) and $|\mathcal{A}_{\alpha,\beta,\gamma}|=0$ otherwise. The whole quantum state involving Alice, Bob, and Charlie is ideally written as $|\Psi(0)\rangle=|\psi\rangle|A_0\rangle|B_0\rangle|C_0\rangle$, where the observers' states are factorized, since they are supposedly locally independent. After the local measurements by Alice, Bob, and Charlie at time $t_1$, but before sharing their information, the quantum state reads: 
\begin{eqnarray}
|\Psi(t_1)\rangle=\sum_{\alpha,\beta,\gamma} \mathcal{A}_{\alpha,\beta,\gamma}| A_{\alpha_x}\rangle|B_{\beta_x}\rangle |C_{\gamma_x} \rangle
\end{eqnarray}   where the observers are now aware of local information concerning the spin-$\frac{1}{2}$ of one particle. {In 
 order to simplify the notation, the degrees  of freedom of the measured particles have been absorbed in the observers' states.} After sharing  their information, Alice, Bob, and Charlie can finally compare their data at time $t_2$:
  \begin{eqnarray}
|\Psi(t_2)\rangle=\sum_{\alpha,\beta,\gamma} \mathcal{A}_{\alpha,\beta,\gamma}| A_{\alpha_x,\beta_x,\gamma_x},B_{\alpha_x,\beta_x,\gamma_x},C_{\alpha_x,\beta_x,\gamma_x}\rangle.
\end{eqnarray} 

 In the latest stage, at time $t_2$, we have four possible orthogonal worlds corresponding to the different outcomes. {We 
 stress that, in general, we do not have factorizability $| A_{\alpha_x,\beta_x,\gamma_x},B_{\alpha_x,\beta_x,\gamma_x},C_{\alpha_x,\beta_x,\gamma_x}\rangle$ $=| A_{\alpha_x,\beta_x,\gamma_x}\rangle|B_{\alpha_x,\beta_x,\gamma_x}\rangle|C_{\alpha_x,\beta_x,\gamma_x}\rangle$ due to entanglement (decoherence)}. However, at time $t_1$, it would be wrong to speak about four worlds in Everett's theory; indeed, each observer is only aware of his/her local result and, for Alice for example, there are only two worlds $| A_{+1_x}\rangle$, $| A_{-1_x}\rangle$ with the same probability $\mathcal{P}_1(\pm 1|\hat{\mathbf{x}},\psi)=\frac{1}{2}$ (the same is true for the other observers).  Moreover, Alice, Bob, and Charlie could have decided to realize different experiments, for example, corresponding to the Stern--Gerlach settings $\hat{\mathbf{n}}_1=\hat{\mathbf{x}}$ and $\hat{\mathbf{n}}_2=\hat{\mathbf{n}}_3=\hat{\mathbf{y}}$. With similar reasoning, the whole quantum state at time $t_1$ and $t_2$ now reads:
      \begin{eqnarray}
|\Psi'(t_1)\rangle=\sum_{\alpha,\beta,\gamma} \mathcal{B}_{\alpha,\beta,\gamma}| A_{\alpha_x}\rangle|B_{\beta_y}\rangle |C_{\gamma_y} \rangle\nonumber\\
|\Psi'(t_2)\rangle=\sum_{\alpha,\beta,\gamma} \mathcal{B}_{\alpha,\beta,\gamma}| A_{\alpha_x,\beta_y,\gamma_y},B_{\alpha_x,\beta_y,\gamma_y},C_{\alpha_x,\beta_y,\gamma_y}\rangle,
\end{eqnarray} with $|\mathcal{B}_{\alpha,\beta,\gamma}|^2=1/4$ iff $\alpha\beta\gamma=+1$ (see Equation~(7)) and $|\mathcal{B}_{\alpha,\beta,\gamma}|=0$ otherwise. In this alternative experiment, we again have at time $t_2$ four worlds with identical probabilities $1/4$. At time $t_1$, Alice (and similarly for Bob and Charlie) has only two worlds $| A_{+1_x}\rangle$, $| A_{-1_x}\rangle$ with the same probability $\mathcal{P}_1(\pm 1|\mathbf{\hat{x}},\psi)=\frac{1}{2}$ as in the previous experiment. This  probability is independent of the measurement choices made by Bob and Charlie, and it gives an illustration of non-signaling that is often used by advocates of Everett's theory to justify their claim for locality.  Again, however, this is the weaker local commutativity that is considered here (using the Schr\"odinger representation) and not local-causality  \`a la Bell. {A 
 similar claim was also formulated  in the Heisenberg and interaction representations by Deutsch and Hayden using the concept of `descriptors' \cite{Deutsch1999,Rubin2001,Horsman,Brassard,Bedard2021}. The conclusions we obtain are of course identical, since the  Schr\"odinger and Heisenberg/interaction pictures are equivalent.}  Therefore, the claims that it could have an impact on the conclusions obtained from Bell--GHZ theorems about nonlocality are clearly unfounded. The mistake here is to believe  that locality defined by local commutativity is the only notion of locality that could be useful or could make sense for Everettians. But as we saw in the answer to \textbf{(A)}, discussing local-causality in the context of Everett's theory makes perfect sense. \\
\indent Often, a  different version of the same attempt to save locality in Everett's theory can be found; it is based on a narrative discussed perhaps for the first time by Page \cite{Page1982} and  Albert and Loewer (in the context of the stochastic many-minds theory) \cite{Albert1988} and reproduced several times since \cite{Tipler2014,Timpson2003,Timpson2015,Deutsch1999,Rubin2001,Horsman,Brassard,Bruce}. The idea is that Alice, Bob, and Charlie  have independently  and locally split into two independent worlds corresponding to their two possible outcomes.  These worlds are independent (decohered) and cannot communicate. It is only when they share their information, which can only occur in the future of the three measurements at the intersection of light cones, that a fourth observer, Dagmara, can correlate the measurements of Alice, Bob, and Charlie. For Everettians, everything is local in this scenario. The splitting is described by the partial density matrices relying only on local-commutativity. The propagation  and sharing of information is also local and made at a velocity that cannot exceed the speed of light. The definition of worlds used here is apparently strongly local but is based only on local-commutativity, not on local-causality.  Of course, when the information from the three measurements arrives at Dagmara, one needs a rule in order to correctly combine the worlds or information about these worlds. This set of rules must depend on the quantum state $\psi$ and on the settings $\hat{\mathbf{n}}_1,\hat{\mathbf{n}}_2,\hat{\mathbf{n}}_3$. For example, following Equation~(2), if Alice measured $|+_x\rangle^1$, the only possibilities allowed by the GHZ quantum state is that Bob and Charlie should have measured either ($|+_x\rangle^2$,$|-_x\rangle^3$) or ($|-_x\rangle^2$,$|+_x\rangle^3$). Alternatively, if Alice measured $|-_x\rangle^1$  we have either  ($|+_x\rangle^2$,$|+_x\rangle^3$) or ($|-_x\rangle^2$,$|-_x\rangle^3$), and other combinations are forbidden by Equation~(2). The strong correlations require an explanation. However, it is at that stage  that the Everettians  are making a mistake. Indeed, once again, it is the deduction surrounding beables $\lambda$ and local-causality that is central here to discuss causality and not local commutativity. Going back to Equations~(14), (17), and (18), we see that what the Everettians are trying is an equation for joint probabilities by Alice, Bob, Charlie, and Dagmara reading as:
\begin{eqnarray}
\mathcal{P}(\alpha,\beta,\gamma|\hat{\mathbf{n}}_1,\hat{\mathbf{n}}_2,\hat{\mathbf{n}}_3,\psi)=\int_\Lambda d\lambda\mathcal{P}_1(\alpha|\lambda,\hat{\mathbf{n}}_1,\psi)\mathcal{P}_2(\beta|\lambda,\hat{\mathbf{n}}_2,\psi)\nonumber\\ 
\times\mathcal{P}_3(\gamma|\lambda,\hat{\mathbf{n}}_3,\psi)
\rho(\lambda|\hat{\mathbf{n}}_1,\hat{\mathbf{n}}_2,\hat{\mathbf{n}}_3,\psi).\nonumber\\
\end{eqnarray} 

In this formula, the three  probabilities $\mathcal{P}_1(\alpha|\lambda,\hat{\mathbf{n}}_1,\psi)$, $\mathcal{P}_2(\beta|\lambda,\hat{\mathbf{n}}_2,\psi)$, $\mathcal{P}_3(\gamma|\lambda,\hat{\mathbf{n}}_3,\psi)$ are obeying the `classical' locality condition of Equation~\eqref{cond1}. However, the density of beables $\rho(\lambda|\hat{\mathbf{n}}_1,\hat{\mathbf{n}}_2,\hat{\mathbf{n}}_3,\psi)$ must depend on the settings $\hat{\mathbf{n}}_1,\hat{\mathbf{n}}_2,\hat{\mathbf{n}}_3$ violating Equation~\eqref{cond2} associated with the absence of superdeterminism. It is in that way that Bell--GHZ theorems are satisfied in the Everettian framework. This is the only causal explanation Everettian proponents of that narrative could propose to justify the strong quantum correlations imposed by the GHZ state. This is the only way of explaining how, for each GHZ game, corresponding to a choice of $\hat{\mathbf{n}}_1,\hat{\mathbf{n}}_2,\hat{\mathbf{n}}_3$, given by Equations (2)--(5), only four possible triplets of outcomes are actualized and picked from the $2^3$ possibilities or worlds obtained, assuming complete independence of the three measurements.  Of course, since there is superdeterminism, this means that  there is no need for a new hidden  `single outcome' assumption here. The crux is the explanation of Bell--GHZ correlations, which requires abandoning local-causality \`a la Bell  (i.e., Equation~\eqref{cond1} or Equation~\eqref{cond2}) in order to recover quantum predictions. Certainly, a model using  Equation~(33)  and relaxing statistical independence can satisfy Bell's theorem (for an example, see the Appendix \ref{Ap2} inspired by \cite{Argaman2010,Drezetarxiv} obtained for the EPR singlet state), but in turn, the conclusion is, once more, that quantum mechanics is nonlocally causal.
 
\indent Interestingly, Everettians sometimes realize that we can define different notions of locality in their theory, but they are getting confused. For instance, David Wallace  introduces the distinction between action-at-a-distance (identified with Bell nonlocality) and non-separability (identified with a form of entanglement). More precisely, for him, ``action-at-a-distance occurs  
when a disturbance to A causes an immediate change in the state of B, without any intervening dynamical process connecting A and B'' \cite{WallaceBook}. This is related to hidden variables as in the de~Broglie--Bohm theory with instantaneous forces \cite{Bohm}. On the other hand, ``nonseparability is a matter not of dynamics but of ontology. A theory is non separable if [...] a complete specification of the state of A and B separately fails to fix the state of the combined system A+B'' \cite{WallaceBook}. This is clearly akin to the mathematical notion of entanglement  used in quantum information where  we can define the measure of entanglement or nonfactorizability (e.g., if the partial density matrix $\hat{\rho}_A=Tr_B[\hat{\rho}_{A,B}]$ for the subsystem A corresponds to a statistical mixture). In the context of Everett's theory where everything is related to the quantum state defining the sole ontology, we are therefore allowed to  speak of \textit{nonlocal facts}. However, Wallace subsequently wrote:
\begin{quote}
``But in Everettian quantum mechanics, violations of Bell's inequality are relatively uninteresting. For Bell's theorem, though its conclusion arguably entails not only non-separability but action-at-a-distance, simply does not apply to the Everett interpretation. It assumes, tacitly, among its premises that experiments have unique, definite outcomes.''~\cite{WallaceBook}
\end{quote} 

Clearly, the explanation  here is related to point \textbf{(A)}: action-at-a-distance (identified with Bell nonlocality) is prohibited because the single outcome  axiom does not apply. However, as we  explained, this is not true; Bell--GHZ theorems do not require such an axiom  and, as we concluded, the many-worlds theory is clearly nonlocally-causal in the sense of Bell. In the same text \cite{WallaceBook}, Wallace  asks: ``Does Everettian quantum mechanics display action-at-a-distance?''   The answer he gave is: ``No. In a quantum field, the quantum state of any region depends only on the quantum state of some cross-section of the past light cone of that region'' \cite{WallaceBook}.  As is clear from the rest of his text, he is now speaking about consequences of local-commutativity and nonsignaling, which is point \textbf{(B)} discussed above. As we explained, local-commutativity cannot be used to conclude the  existence  of Bell nonlocality.  Therefore, the slight but still visible change in the definition of locality  results in even more confusion between \textbf{(A)}  and \textbf{(B)}. In the end, Everettians are probably ready to accept non-separability as a form of nonlocality. However, the difference with Bell nonlocality is not clearly stated by Everettians. This is particularly problematic since it is known, at least for pure quantum states $|\Psi\rangle$ that form the sole ontology of Everett's theory, that it is always possible to define measurements inducing violations of some Bell's inequalities~\cite{Liang}.\\
\indent The whole issue is clearly related to the complex interplay between ontic and epistemic aspects  of spatio-temporal facts  in the many-worlds theory. Indeed, returning to the important works of Wallace (see Figures 8.1 and 8.2 in Chapter~8 of \cite{WallaceBook}) and Carroll (see Figure p.~171 in Chapter~8 of \cite{Carroll}), we see that there is an  ambiguity or perhaps underdetermination concerning the nature and location of wave function splitting in Everett's theory. If Alice, Bob, and Charlie separately measure the spin of their particles, some advocates of the many-worlds interpretation would prefer to think that, before sharing the information with Dagmara (i.e., at the intersection of their future light cones), it is not even definite to speak about one single common reality. This is why, they say, the theory is local.  This is clearly the perspective taken by Brassard and Raymond-Robichaud \cite{Brassard} and agrees with the Deutsch--Hayden descriptor approach in the Heisenberg representation~\mbox{\cite{Deutsch1999,Rubin2001,Horsman,Brassard,Bedard2021}}. A similar view is also shared by Brown and Timpson, who wrote:
\begin{quote}
``we can only think of the correlations between measurement outcomes on the two sides of the experiment actually obtaining in the overlap of the future light-cones of the measurement events---they do not obtain before then and---a fortiori---they do not obtain instantaneously.'' \cite{Timpson2015}
\end{quote}

This conclusion is strongly related to our Equation (30) for $|\Psi(t_1)\rangle$ at time $t_1$ before Alice, Bob, and Charlie shared their information. Indeed, $|\Psi(t_1)\rangle$ is split into four orthogonal terms and, from the point of view of the ontology of the many-worlds theory where everything is based on the quantum state, it seems that we are justified to identify four `worlds'. Moreover, the information {\em locally} available to each observer is different, since we can always rewrite the state $|\Psi(t_1)\rangle$ so that Alice can only be in two `worlds' (the same is separately true for Bob and Charlie).  Moreover, the word local used here is certainly not the same as the one used by Bell in defining local-causality. What Brown and Timpson \cite{Timpson2015} discuss is actually reduced information obtained by observers  (agents) and can be summarized by the reduced density matrices. In other words, they are talking about non-signaling fully justified by local-commutativity. This is therefore different from Bell-locality involving causality and beables $\lambda$. Interestingly, Vaidman \cite{Vaidman2015,Vaidman2012}, together with McQueen \cite{McQueen2019}, discussed this problem in order to justify Born's rule in the many-worlds interpretation based on locality and symmetry. Their analysis  actually involves a discussion of the information available to the different observers but as recognized by Vaidman long ago:
\begin{quote}
``The above examples illustrate how the MWI clarifies the issue of nonlocality in quantum mechanics. The quantum state of the universe, and its components which correspond to various worlds, are nonlocal. Interactions governed by the Hamiltonian evolution change locally the state of the universe. [...] Thus, in the created worlds we obtain, effectively, nonlocal changes, while there is no nonlocal action on the physical level of the universe.''~\cite{Vaidman1994}
\end{quote}  

In other words, Everett's theory can be both local and nonlocal, and this means that we must carefully distinguish the different meanings of the word local. What we showed in this work  is that the concept of local-causality  introduced by Bell and involving beables  $\lambda$ (i.e., causes or explanations) can be rigorously discussed even in the framework of the many-worlds interpretation. When we do that, unsurprisingly, Everett's theory appears non locally-causal. Still, we can discuss the problem of local information available by individual observers and introduce  non-signaling  using  only local-commutativity. We can also introduce  quantum entanglement (i.e., correlation), and correlations are not causes! The different perspectives are not contradictory if we clearly distinguish the  nonlocal ontology from the local information available to agents.       
\section{Conclusions: Symmetry, Democracy, and Nonlocality}\label{Sec4}
\indent A last point that we avoided until now and that nevertheless plays a role in the discussion is the problem of defining probabilities in Everett's theory. The GHZ theorem was used by Vaidman~\cite{Vaidman2015} mostly in order to avoid this problem. Indeed, with  the GHZ state, what is needed is the definition of probability $\mathcal{P}=0$ or 1 associated with certainty. This is what allowed us, through Equations (21)--(23), to derive determinism and subsequently obtain the logical contradiction leading to the unavoidable conclusion: quantum mechanics is nonlocally-causal.\\
\indent However, in  order to be quantitative, one must go beyond certainty, and this clearly requires some Bell inequalities involving probabilities different from 0 or 1 (this is central if detectors with low efficiencies are taken into account). However, it is well-known that defining probabilities and therefore recovering quantum statistical predictions is very difficult for Everettians.  Everett himself, starting from the fact that a pure unitary theory does not contains stochastic elements, originally called his theory quantum mechanics without  probability \cite{Barrett2012}.  He instead introduced a `typicality' measure $\mathcal{M}$ that, formally speaking, is the same as Born's probability  squaring wave function amplitudes (for example, in Equation (30) $\mathcal{M}_{\alpha,\beta,\gamma}:=|\mathcal{A}_{\alpha,\beta,\gamma}|^2$). Using the weak law of large numbers, he was able to show that, in the long run (i.e., after repeating the same quantum experiment many times), the typicality measure $\mathcal{M}$ of having deviations between the frequency of occurence $f_\alpha$ (of any observable event $\alpha$) and the measure  $\mathcal{M}_\alpha$ of this  particular event tends to vanish:
\begin{eqnarray}
\lim_{N\rightarrow +\infty}\mathcal{M}(\{|f_\alpha-\mathcal{M}_\alpha|>\epsilon_\alpha\})=0,
\end{eqnarray}  with $\epsilon_\alpha>0$ and $N$ being the number of repetitions. The problem, as has been emphasized many times, is that such a proof of equality between observed frequency $f_\alpha$ and measure $\mathcal{M}_\alpha$ is circular.  This is already very well-known in the language of probability with Bernoulli processes,  and the new semantic using `typicality measure' does not change anything about the situation. As was correctly concluded by Wallace:\begin{quote} ``We cannot prove that, in the long-run, relative frequencies converge to probabilities. What we can prove, is that in the long-run relative frequencies converge to probabilities ... probably''~\cite{Wallacevideo}.
\end{quote} 

Morphing probability into a typicality measure does not help, and this `proof' has been strongly criticized over the years, in particular because it relies on many subtleties concerning the application of the Bernoulli law of large-numbers, e.g.,  concerning infinite sequences of repeated experiments (for reviews and discussions, see \cite{Barrett2017,Drezetarxiv2021}). Perhaps the key problem here is that  Everettian theory is actually a purely deterministic field theory and,  in such a theory, all branches or alternatives have the same importance. Defining a probability--typicality measure using Born's rule is allowed, but this breaks the symmetry or democracy between the branches; the theory does not contain enough ingredients to unequivocally fix the problem~\cite{Drezetarxiv2021}. There are infinite ways for defining probabilities, and the theory cannot decide because there is no way to decide!  In recent years, the mere idea that a deterministic ontology simply based on Schr\"odinger's unitary equation could lead to something like a justification of the probabilistic properties of our universe has been named the `incoherence problem' by philosophers~\cite{Albert2015,Maudlin2019}.   Nevertheless, it has  been claimed in reply to criticisms that probabilistic concepts used by Everettians are not worse (and perhaps not better) than they are in other interpretations of quantum mechanics or even in classical statistical mechanics \cite{WallaceBook}. In parallel, Vaidman~\cite{Vaidman2020}, partly in response to Albert~\cite{Albert2015}, has developed suggestive narratives and 	a complete semantics for speaking about probability and `self-locating uncertainty' as understood and perceived by observers in quantum branches of the universal wavefunction. Morever, to add to the situation, even more confusing several new ideas based on decision-theoretic scenarios  \`a  la Deutsch~\cite{Deutsch1999b,WallaceBook} or `envariance' \`a la Zurek \cite{Zurek2003,Zurek2005} have been discussed as alternative `proofs' for making sense of probability in the many-worlds interpretation, i.e.,  without incoherence and with the correct Born's~rule.\\
\indent   All these new discussions and methods are certainly interesting, but there is no consensus, and we can think that they do not really solve or clarify the issue about probabilities and Born's rule in Everett's theory. Indeed, in the end, what we obtain are only self-consistent rules based on some other principles and the fact that if one could physically and mathematically define a  probability in the many-worlds interpretation, then this probability  should naturally be given by Born's rule. Of course, the fundamental problem remains: are we right in assigning probabilities to events in the many-worlds interpretation? A clear answer is  still missing,  we think this is impossible without adding new physical axioms foreign to the pure unitary theory of Everett.\\
\indent An example of such an attempt is the many-minds model we proposed recently~\cite{DrezetMWI}, but this relies on very speculative ideas about minds  and initial conditions in a quantum universe.  A more serious approach is perhaps the many-Bohmian paths interpretation that has been introduced by several authors~\cite{Valentini2010,Tipler2014,Bostrom2014,Sebens2014,Hall2014,Tappenden}. In this approach, Bohmian mechanics, i.e., an hidden-variables theory, is fundamental, and we can define a `multiverse'  by taking a very large universe or set, including many separated (and independent) copies of the same subsystem (e.g., our local universe). Each subsystem is described by the same wave function $\Psi$ (up to some translation in space), and the multiverse is characterized by a wave function product $\otimes \Psi$.  Since we have a very large ensemble or `collective' of universes, and since Bohmian mechanics requires additional initial conditions for the particles or other field beables defining the ontology,  we can apply the law of large numbers to this theory by postulating by fiat, i.e., as an initial condition of the whole multiverse,  that the probability--typicality measure for these beables is given by Born's rule.    In the de~Broglie--Bohm or Bohmian mechanics \cite{Sheldon}, this makes perfect sense, and therefore we can, in the long run (i.e., if the number of copies of $N$ is very large), recover quantum statistical predictions.\\
\indent  Moreover, it is well known that the de~Broglie--Bohm theory is Bell-nonlocal \cite{Bohm}, and this means that the many-Bohmian paths interpretation must be nonlocal as well~\cite{Hall}. This again refutes the locality claims of Everettians. At the end of this analysis, we can fairly conclude that Everett's theory is either  wrong or  nonlocally-causal.       

\section{Appendix }
\subsection{Appendix 1}
\label{Ap0}
\indent A central issue for the many-worlds theory concerns ontology and the status of the wavefunction. In Everett's theory, it is usually assumed that the wave function is the fundamental field or beable of the theory. In an older work \cite{Drezet2016}, we advocated the view that the $\psi-$field in the many-worlds theory must be seen  as a nonlocal generalization of classical field theory such as Maxwell's electromagnetism, which is based on local electric and magnetic  fields $\mathbf{E}(\mathbf{x},t)$ and $\mathbf{B}(\mathbf{x},t)$, defined at a point $\mathbf{x}$ and time $t$. Following the original Schr\"odinger's interpretation, the $\psi-$field $\psi(\mathbf{x},t)$ is here seen as a material field describing the delocalized single electron, and Everett's theory is simply this theory pushed to its logical conclusion for many electrons or many `particles'. However, in this theory (unlike in de~Broglie--Bohm mechanics~\cite{Bohm}), there are no point particles, just a wavefunction $\psi(\mathbf{x}_1,...,\mathbf{x}_N,t)$ that depends on $N$ position vectors $\mathbf{x}_j$ (since $\psi$ is a correlation field, this was identified with a `webfunction' in \cite{Drezet2016} and with a `multi-field' in \cite{Hubert2018}). Alternatively, the wave function can be defined in the 3N-dimensional configuration space, but this is mathematically and therefore physically equivalent. The theory can be generalized to include spins and quantum fields. For example, for the the GHZ state, we can define $\psi_{\alpha,\beta,\gamma}(\mathbf{x}_1,\mathbf{x}_2,\mathbf{x}_3, t)$ where $\alpha,\beta,\gamma=\pm 1$ label spinor components along some directions. Moreover,  $\psi$ is in general a nonlocal object in the sense that it is non-factorizable due to entanglement.      \\
\indent It is, however, possible to define local beables in analogy with Bohmian mechanics. Indeed, in Bohmian mechanics, we can define trajectories for particles using coupled first-order differential equations. For the three particles  of the GHZ state, we have $\frac{d}{dt}\mathbf{x}_j(t)=\mathbf{F}_j(\mathbf{x}_1(t)\mathbf{x}_2(t),\mathbf{x}_3(t),t)$, where the three nonlocal functions $F_j$  \cite{Bohm,Bohm2} characterize a dynamics displaying action-at-a-distance. Moreover, we can use these trajectories as a hydrodynamical (i.e., Madelung) representation of the Schr\"odinger $\psi-$field and therefore define local beables $\lambda\equiv [\mathbf{x}_1(t)\mathbf{x}_2(t),\mathbf{x}_3(t)]$ for Everett's theory. Furthermore, in this hydrodynamical approach, the fluid density $\psi^\dagger\psi$ defines a conserved measure $\mathcal{M}$ which is transported by the trajectories during the time evolution. As explained in Section \ref{Sec4}, $\mathcal{M}$ is not a probability. In order to discuss local-causality in the context of such a theory, it is thus only necessary to consider  some initial data in the remote past and show that the deterministic Bohmian dynamics violates Equations~\eqref{cond1} and \eqref{cond2}, where probabilities $\mathcal{P}$ are replaced by fluid measure $\mathcal{M}$. It is, in particular, possible to show that the elementary fluid measure $d\mathbf{M}(\mathbf{x}_1(t),\mathbf{x}_2(t),\mathbf{x}_3(t),t)$ (defined in infinitesimal comoving volumes surrounding the points $\mathbf{x}_1(t),\mathbf{x}_2(t),\mathbf{x}_3(t)$) is nonlocally affected by operations completed by the three observers Alice, Bob, and Charlie. For this, it is sufficient to consider examples used in Bohmian mechanics where the nonlocal features are already  identified (see, for example, \cite{Bricmont,Drezet2019}).  Moreover, the local measure $d\mathbf{M}_1(\mathbf{x}_1(t),t)$, obtained by Alice after integrating $d\mathbf{M}(\mathbf{x}_1(t),\mathbf{x}_2(t),\mathbf{x}_3(t),t)$ over the  beables $\mathbf{x}_2(t),\mathbf{x}_3(t)$ associated with the Bob and Charlie sides, is clearly independent of any local operations done by Bob and Charlie in agreement  with the nonsignaling theorem discussed in Section \ref{Sec4}.  We stress that in \cite{Allori}, it was similarly shown that Everett's theory can be `locally' formulated in term of mass density.  Moreover, this representation agrees with the definition of local-commutativity in that Bell nonlocal correlations between subsystems are hidden when averaged on the remote variables associated with the other observer. 

\subsection{Appendix 2}
\label{Ap1}
\indent The history of Bell locality is well-documented, either in Bell's own writings  on the subject \cite{Bell} or in several articles deeply analyzing the concept~\cite{Bell2023,Maudlin2014,Norsen2011,Laudisa2023}.  The concept of local-causality was explictly introduced by Bell in 1976 in  `The theory of local beables' (see \cite{Bell}, Chapter~7, pp.~52--62) in connection with the notion of beables $\lambda$, and the concept was further analyzed in `La nouvelle cuisine' (see \cite{Bell}, Chapter~24, pp.~232--248).  Here, the goal is not to discuss these works but simply to connect them  with the GHZ nonlocality derivation~\cite{GHZ}. For this, we return to Equation~(14) and to the conditional probability $\mathcal{P}(\alpha,\beta,\gamma|\lambda,\hat{\mathbf{n}}_1,\hat{\mathbf{n}}_2,\hat{\mathbf{n}}_3,\psi)$. By the definition of conditional  probabilities, we can always~write: 
\begin{eqnarray}
\mathcal{P}(\alpha,\beta,\gamma|\lambda,\hat{\mathbf{n}}_1,\hat{\mathbf{n}}_2,\hat{\mathbf{n}}_3,\psi)=\mathcal{P}(\alpha|\beta,\gamma,\lambda,\hat{\mathbf{n}}_1,\hat{\mathbf{n}}_2,\hat{\mathbf{n}}_3,\psi)\mathcal{P}(\beta|\gamma,\lambda,\hat{\mathbf{n}}_1,\hat{\mathbf{n}}_2,\hat{\mathbf{n}}_3,\psi)\nonumber\\ \times\mathcal{P}(\gamma|\lambda,\hat{\mathbf{n}}_1,\hat{\mathbf{n}}_2,\hat{\mathbf{n}}_3,\psi).
\end{eqnarray} 

Now $\alpha,\beta,\gamma$ are beables recorded by Alice, Bob, and Charlie at three space time points A, B, and C separated by space-like intervals. The Bell-locality postulate \cite{Bell2023,Maudlin2014,Norsen2011,Laudisa2023} imposes that we must have:
  \begin{eqnarray}
\mathcal{P}(\alpha|\beta,\gamma,\lambda,\hat{\mathbf{n}}_1,\hat{\mathbf{n}}_2,\hat{\mathbf{n}}_3,\psi)=\mathcal{P}_1(\alpha|\lambda,\hat{\mathbf{n}}_1,\psi)\\ 
\mathcal{P}(\beta|\gamma,\lambda,\hat{\mathbf{n}}_1,\hat{\mathbf{n}}_2,\hat{\mathbf{n}}_3,\psi)=\mathcal{P}_2(\beta|\lambda,\hat{\mathbf{n}}_2,\psi)\\ 
\mathcal{P}(\gamma|\lambda,\hat{\mathbf{n}}_1,\hat{\mathbf{n}}_2,\hat{\mathbf{n}}_3,\psi)=\mathcal{P}_3(\gamma|\lambda,\hat{\mathbf{n}}_3,\psi).
\end{eqnarray}

These relations (leading to Equation~\eqref{cond1}) presuppose that the conditional probabilities associated with each point A, B, and C are independent of outcomes and parameters associated with the other detection points. This is more rigorously justified if the parameters  $\hat{\mathbf{n}}_1,\hat{\mathbf{n}}_2,\hat{\mathbf{n}}_3$ 	are determined by information that is not coming from the overlap of the three  past light  cones with apexes A, B, and C (for example, the choice of the setting  directions could be influenced by some supposedly independent photons coming from remote galaxies~\cite{Zeilinger}). The local beables $\lambda$ are oppositely contained in the overlap of these three past light cones in order to define a past common cause.\\
\indent Moreover, we stress that Bell implicitly assumed the statistical independence in Equation~\eqref{cond2} $\rho(\lambda|\hat{\mathbf{n}}_1,\hat{\mathbf{n}}_2,\hat{\mathbf{n}}_3,\psi)=\rho(\lambda|\psi)$. This can be rigorously formulated if, once more, $\lambda$ are  contained in the overlap of the three past light cones with apexes A, B, and C, whereas the parameters  $\hat{\mathbf{n}}_1,\hat{\mathbf{n}}_2,\hat{\mathbf{n}}_3$ 	are determined by information that is not coming from the overlap. This shows that Equations~\eqref{cond1} and~\eqref{cond2} rely on some common assumptions about causality and locality, i.e., constrained by the requirement of past common causes and by the geometry of spacetime light cones. This justifies why we can group Equations~\eqref{cond1} and~\eqref{cond2} in order to define a single concept of  Bell-locality. We also stress that, even if  Equation~\eqref{cond2} is often supposed to be a natural prerequisite of science, its validity is, however, questionned by advocates of superdeterministic approaches (for a discussion, see \cite{Chen}). 

\subsection{Appendix 3}
\label{Ap2}
We want to find an explicit representation of Equation~(33) for the GHZ  state. A solution limited to the four kinds of measurements needed for the GHZ nonlocality proof~reads:
\begin{eqnarray}
\mathcal{P}(\alpha,\beta,\gamma|\hat{\mathbf{x}},\hat{\mathbf{x}},\hat{\mathbf{x}},\psi)=\oint  d\theta_1d\theta_2d\theta_3\mathcal{P}_1(\alpha|\theta_1,\hat{\mathbf{x}},\psi)\mathcal{P}_2(\beta|\theta_2,\hat{\mathbf{x}},\psi)\nonumber\\ \times\mathcal{P}_3(\gamma|\theta_3,\hat{\mathbf{x}},\psi)
\rho(\theta_1,\theta_2,\theta_3|\hat{\mathbf{x}},\hat{\mathbf{x}},\hat{\mathbf{x}},\psi)\nonumber\\
\end{eqnarray} \vspace{-6pt}
\begin{eqnarray}
\mathcal{P}(\alpha,\beta,\gamma|\hat{\mathbf{x}},\hat{\mathbf{y}},\hat{\mathbf{y}},\psi)=\oint  d\theta_1d\theta_2d\theta_3\mathcal{P}_1(\alpha|\theta_1,\hat{\mathbf{x}},\psi)\mathcal{P}_2(\beta|\theta_2,\hat{\mathbf{y}},\psi)\nonumber\\ \times\mathcal{P}_3(\gamma|\theta_3,\hat{\mathbf{y}},\psi)
\rho(\theta_1,\theta_2,\theta_3|\hat{\mathbf{x}},\hat{\mathbf{y}},\hat{\mathbf{y}},\psi)\nonumber\\
\end{eqnarray} \vspace{-6pt}
\begin{eqnarray}
\mathcal{P}(\alpha,\beta,\gamma|\hat{\mathbf{y}},\hat{\mathbf{x}},\hat{\mathbf{y}},\psi)=\oint  d\theta_1d\theta_2d\theta_3\mathcal{P}_1(\alpha|\theta_1,\hat{\mathbf{y}},\psi)\mathcal{P}_2(\beta|\theta_2,\hat{\mathbf{x}},\psi)\nonumber\\ \times\mathcal{P}_3(\gamma|\theta_3,\hat{\mathbf{y}},\psi)
\rho(\theta_1,\theta_2,\theta_3|\hat{\mathbf{y}},\hat{\mathbf{x}},\hat{\mathbf{y}},\psi)\nonumber\\
\end{eqnarray} \vspace{-6pt}
\begin{eqnarray}
\mathcal{P}(\alpha,\beta,\gamma|\hat{\mathbf{y}},\hat{\mathbf{y}},\hat{\mathbf{x}},\psi)=\oint  d\theta_1d\theta_2d\theta_3\mathcal{P}_1(\alpha|\theta_1,\hat{\mathbf{y}},\psi)\mathcal{P}_2(\beta|\theta_2,\hat{\mathbf{y}},\psi)\nonumber\\ \times\mathcal{P}_3(\gamma|\theta_3,\hat{\mathbf{x}},\psi)
\rho(\theta_1,\theta_2,\theta_3|\hat{\mathbf{y}},\hat{\mathbf{y}},\hat{\mathbf{x}},\psi)\nonumber\\
\end{eqnarray} 
 with the local conditional probabilities: 
 \begin{eqnarray}
 \mathcal{P}_j(\alpha|\theta_j,\hat{\mathbf{x}},\psi)=\frac{1+\alpha\cos{(\theta_j)}}{2}\nonumber\\
 \mathcal{P}_j(\alpha|\theta_j,\hat{\mathbf{y}},\psi)=\frac{1+\alpha\sin{(\theta_j)}}{2} & j=1,2,3
 \end{eqnarray} 
 and the four different superdeterministic probability densities:
  \begin{eqnarray}
  \rho(\theta_1,\theta_2,\theta_3|\hat{\mathbf{x}},\hat{\mathbf{x}},\hat{\mathbf{x}},\psi)=\frac{1}{4}\sum_{\substack{v_1,v_2,v_3 =\pm 1\\ 
  \textrm{iff }v_1\cdot v_2\cdot v_3=-1}}\delta(\hat{\boldsymbol{\theta}}_1-v_1\hat{\mathbf{x}})\delta(\hat{\boldsymbol{\theta}}_2-v_2\hat{\mathbf{x}})\delta(\hat{\boldsymbol{\theta}}_3-v_3\hat{\mathbf{x}})\nonumber\\
  \rho(\theta_1,\theta_2,\theta_3|\hat{\mathbf{x}},\hat{\mathbf{y}},\hat{\mathbf{y}},\psi)=\frac{1}{4}\sum_{\substack{v_1,v_2,v_3= \pm 1 \\ 
  \textrm{iff }v_1\cdot v_2\cdot v_3=+1}}\delta(\hat{\boldsymbol{\theta}}_1-v_1\hat{\mathbf{x}})\delta(\hat{\boldsymbol{\theta}}_2-v_2\hat{\mathbf{y}})\delta(\hat{\boldsymbol{\theta}}_3-v_3\hat{\mathbf{y}})\nonumber\\
  \rho(\theta_1,\theta_2,\theta_3|\hat{\mathbf{y}},\hat{\mathbf{x}},\hat{\mathbf{y}},\psi)=\frac{1}{4}\sum_{\substack{v_1,v_2,v_3=\pm 1 \\ 
  \textrm{iff }v_1\cdot v_2\cdot v_3=+1}}\delta(\hat{\boldsymbol{\theta}}_1-v_1\hat{\mathbf{y}})\delta(\hat{\boldsymbol{\theta}}_2-v_2\hat{\mathbf{x}})\delta(\hat{\boldsymbol{\theta}}_3-v_3\hat{\mathbf{y}})\nonumber\\
  \rho(\theta_1,\theta_2,\theta_3|\hat{\mathbf{y}},\hat{\mathbf{y}},\hat{\mathbf{x}},\psi)=\frac{1}{4}\sum_{\substack{v_1,v_2,v_3=\pm 1 \\ 
  \textrm{iff }v_1\cdot v_2\cdot v_3=+1}}\delta(\hat{\boldsymbol{\theta}}_1-v_1\hat{\mathbf{y}})\delta(\hat{\boldsymbol{\theta}}_2-v_2\hat{\mathbf{y}})\delta(\hat{\boldsymbol{\theta}}_3-v_3\hat{\mathbf{x}})\nonumber\\
  \end{eqnarray} involving the definitions:
  \begin{eqnarray}
  \delta(\hat{\boldsymbol{\theta}}_j-\hat{\mathbf{x}})\equiv \delta(\theta_j),&  \delta(\hat{\boldsymbol{\theta}}_j+\hat{\mathbf{x}})\equiv \delta(\theta_j-\pi)\nonumber\\
  \delta(\hat{\boldsymbol{\theta}}_j-\hat{\mathbf{y}})\equiv \delta(\theta_j-\frac{\pi}{2}),&  \delta(\hat{\boldsymbol{\theta}}_j+\hat{\mathbf{y}})\equiv \delta(\theta_j+\frac{\pi}{2}).
  \end{eqnarray} 

\end{document}